\documentclass[dvips,12pt,aps,prc,reprint,groupedaddress,superscriptaddress,showpacs]{revtex4-1}
\usepackage{natbib,lipsum}
\usepackage{graphicx,subfigure,float,graphics,amssymb,amsmath,color}
\usepackage{epsfig}

\begin{document}

\title{Semiclassical description of chiral geometry in triaxial nuclei}

\author{R. Budaca}
\affiliation{"Horia Hulubei" National Institute for Physics and Nuclear Engineering, Str. Reactorului 30, RO-077125, POB-MG6 Bucharest-M\v{a}gurele, Romania}
\affiliation{Academy of Romanian Scientists, 54 Splaiul Independen\c{t}ei, RO-050094, Bucharest, Romania}

\date{\today}

\begin{abstract}
A triaxial particle-rotor Hamiltonian for three mutually perpendicular angular momentum vectors corresponding to two high-$j$ quasiparticles and the rotation of a triaxial collective core, is treated within a time-dependent variational principle. The resulting classical energy function is used to investigate the rotational dynamics of the system. It is found that the classical energy function exhibits two minima starting from a critical angular momentum value which depends on the single-particle configuration and the asymmetry measure $\gamma$. The emergence of the two minima is attributed to the breaking of the chiral symmetry. Quantizing the energy function for a given angular momentum, one obtains a Schr\"{o}dinger equation with a coordinate dependent mass term for a symmetrical potential which changes from a single to a double well shape as the angular momentum pass the critical value. The energies of the chiral partner bands for a given angular momentum are then given by the lowest two eigenvalues. The procedure is exemplified for maximal triaxiality and two $h_{11/2}$ quasiparticles, with the results used for the description of the chiral doublet bands in $^{134}$Pr.
\end{abstract}


\maketitle

\section{Introduction}

The concept of chirality or handedness is a common occurrence in biology, chemistry, optics, and particle physics. In nuclear physics, chirality is associated with the geometry of three mutually perpendicular angular momenta. It was originally suggested by Frauendorf and Meng \cite{FrauMeng} for a system composed of a triaxial core coupled to a set of high-$j$ valence particles and holes. The rationale for this particular ensemble \cite{FrauMeng,Meng1} is that the triaxial core tends to rotate around the axis with the largest moment of inertia (MOI) which imply an intermediate density distribution, while the motion of particles and holes prefer ellipsoidal orbits following maximal and respectively minimal density distributions around the other two principal axes. The three mutually perpendicular angular momenta form a screw in respect to the total angular momentum vector and therefore can be arranged to form two systems with opposite intrinsic chirality. As the broken chiral symmetry should be restored in the laboratory frame of reference, one expects to observe two nearly degenerate $\Delta I=1$ bands with the same parity. This particular signature, {\it i.e.} the so-called chiral doublet bands, was first observed in few $N=75$ odd-odd isotones \cite{Starosta}. The experimental confirmation of chiral symmetry breaking was followed by an extensive search for other candidate nuclei. Such that, presently chiral bands are reported in over 30 nuclei clustered in "\emph{islands}" of chirality around mass numbers 80, 100, 130 and 190, where the chiral geometry is generated by specific quasiparticle configurations \cite{Bark,StarKoike}.

The original interpretation of chirality \cite{FrauMeng} was based on both the particle-rotor (PRM) \cite{BM} and tilted axis cranking models (TAC) \cite{Tac}. Alternative descriptions of the chiral bands include presently boson expansion approaches \cite{Brant,Ganev,Raduta1,Raduta2}, and Shell Model based formalisms \cite{SM1,SM2}. However, being a fully quantum model, and therefore capable of treating the tunneling between the two chiral solutions, PRM remained the standard for theoretical studies of chirality \cite{PRM1,PRM2,PRM3,PRM4,PRM5,PRM6,PRM7,PRM8,PRM9,PRM10}. Although the semiclassical nature of the cranking mean field approaches is not able to describe the quantum interaction between the chiral bands, it has the advantage of providing a relation between the density distribution and the direction of the total angular momentum vector \cite{TAC1,TAC2,TAC3,TAC4,TAC5,TAC6,TAC7}. Moreover it can be easily extrapolated to multi-quasiparticle configurations, whereas the PRM advances in this direction are incipient \cite{MultiPRM1,MultiPRM2,MultiPRM3}. The need of going beyond mean field approximation produced successful extensions of the TAC formalism such as the TAC plus random phase approximation \cite{TACRPA1,TACRPA2}, and the collective Hamiltonian approach \cite{col1,col2}. The latter take advantage of the information on classical rotational dynamics obtained from TAC calculations to construct a quantum collective Hamiltonian, whose solutions were shown to be close to the fully quantum and exact PRM calculations.

In this paper, one will take the opposite approach in combining the advantages of the classical and quantum pictures by treating semiclassically a particle-rotor type Hamiltonian. The semiclassical procedure amounts to ascribing a time-dependent variational principle to the quantum Hamiltonian, which is consequently dequantized into a classical energy function. A similar procedure was already successfully applied for the description of wobbling excitations in odd mass nuclei \cite{Raduta,Budaca1}. By choosing an appropriate variational function one can select a limited set of degrees of freedom relevant for the studied phenomenon instead of treating the full space. The information on the rotational dynamics of the system is then extracted from the evolution of the classical energy function as well as other observables expressed in terms of the azimuthal and polar angles with the variation of the total angular momentum which retains its quality of good quantum number. The emergence of chiral solutions at a certain spin is discussed from the classical point of view. For the description of the chiral partner bands, the classical energy function is quantized in respect to a chiral variable. The similarities and the differences between the resulting Schr\"{o}dinger equation and the chiral collective Hamiltonian of Refs.\cite{col1,col2} are pointed out. The formalism is applied to the description of the chiral bands in $^{134}$Pr.

\setcounter{equation}{0}
\renewcommand{\theequation}{2.\arabic{equation}}
\section{Semiclassical approach}

The following extension of the particle-rotor Hamiltonian \cite{BM}
\begin{equation}
H=H_{R}+H_{sp}+H'_{sp},
\end{equation}
is employed for the description of the interaction between two single-particle angular momenta and a collective one. $H_{R}=\sum_{k=1,2,3}A_{k}(\hat{I}_{k}-\hat{j}_{k}-\hat{j'}_{k})^{2}$ is the triaxial rotor Hamiltonian associated to the core angular momentum $\vec{R}=\vec{I}-\vec{j}-\vec{j'}$ and defined by the inertial parameters $A_{k}=1/(2\mathcal{J}_{k})$ where $\mathcal{J}_{k}$ are the MOI along the principal axes of the intrinsic frame of reference considered in the hydrodynamic estimation \cite{BM}:
\begin{equation}
\mathcal{J}_{k}=\frac{4}{3}\mathcal{J}_{0}\sin^{2}{\left(\gamma-\frac{2}{3}k\pi\right)}.
\label{inert}
\end{equation}
$\vec{j}$ and $\vec{j'}$ are single-particle generated spins, {\it i.e.} they can be the total angular momentum of a single quasiparticle orbital or a resultant spin of few quasiparticles. The single-particle contribution to the total Hamiltonian coming from single-particle spin $\vec{j}$ is
\begin{eqnarray}
H_{sp}&=&\frac{V}{j(j+1)}\left\{\left[3\hat{j}_{3}^{2}-j(j+1)\right]\cos{\gamma}\right.\nonumber\\
&&\left.-\sqrt{3}(\hat{j}_{1}^{2}-\hat{j}_{2}^{2})\sin{\gamma}\right\},
\end{eqnarray}
where $\gamma$ is the asymmetry parameter, which also defines the ratios between MOI.

\begin{figure}[t!]
\begin{center}
\includegraphics[width=0.41\textwidth]{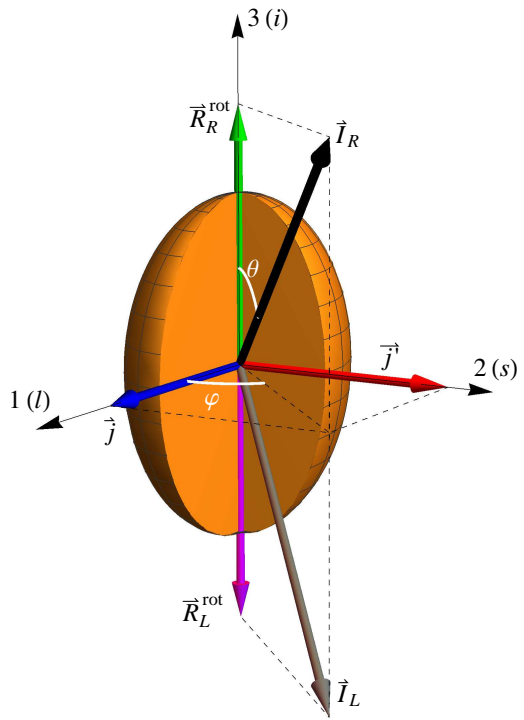}
\end{center}
\caption{Schematic representation of the chiral geometry.}
\label{elipsoid}
\end{figure}

Suppose that each of the single-particle angular momenta is aligned to a principal axis of the intrinsic frame of reference. Although the angular momentum of a triaxial rotor is actually distributed on all three axes, the core will rotate around the axis with the highest MOI, which is then chosen as a quantization axis. As the absolute value of the rotor spin increases, the contributions from the other two axes become smaller \cite{Shi} and can be quantized for example into wobbling excitations \cite{BM}. If this rotation axis is perpendicular to the plane of the two single-particle spins, then one will have a trihedral vector configuration as in Fig.\ref{elipsoid}. In order to have the highest MOI along the third intrinsic axis, $\gamma$ must be within the interval $(60^{\circ},120^{\circ})$. In this $\gamma$ interval, the ellipsoid's semi-axes
\begin{equation}
R_{k}=R_{0}\left[1+\sqrt{\frac{5}{4\pi}}\beta\cos{\left(\gamma-\frac{2\pi}{3}k\right)}\right]
\end{equation}
are arranged as $R_{2}<R_{3}<R_{1}$. In order to keep track of the direction of the total angular momentum vector relative to the density distribution, the axes 1,2 and 3 are also referred to as the long ($l$), short ($s$) and medium ($i$).

Choosing the single-particle alignment to be rigid along the axes 1 and 2, i.e. $\hat{j}_{1}\approx j\equiv const.$ and $\hat{j'}_{2}\approx j'\equiv const.$, the Hamiltonian relevant for the system's dynamics can be limited to:
\begin{equation}
H=A_{1}(\hat{I}_{1}-j)^{2}+A_{2}(\hat{I}_{2}-j')^{2}+A_{3}\hat{I}_{3}^{2}+const.
\end{equation}
Thus, the Hamiltonian to be treated is:
\begin{equation}
H_{chiral}=A_{1}\hat{I}_{1}^{2}+A_{2}\hat{I}_{2}^{2}+A_{3}\hat{I}_{3}^{2}-2A_{1}j\hat{I}_{1}-2A_{2}j'\hat{I}_{2}.
\label{Hal}
\end{equation}

For the purpose of investigating the rotational motion described by the quantum Hamiltonian (\ref{Hal}), one considers the variational principle
\begin{equation}
\delta\int_{0}^{t}\langle\psi(z)|H_{chiral}-\frac{\partial}{\partial{t'}}|\psi(z)\rangle dt'=0.
\label{var}
\end{equation}
The variational state is chosen of the form:
\begin{eqnarray}
|\psi(z)\rangle&=&\sum_{K=-I}^{I}\sqrt{\frac{(2I)!}{(I-K)!(I+K)!}}\frac{z^{I+K}}{\left(1+|z|^{2}\right)^{I}}|IMK\rangle\nonumber\\
&=&\frac{1}{\left(1+|z|^{2}\right)^{I}}e^{z\hat{I}_{-}}|IMI\rangle.
\label{coh}
\end{eqnarray}
This is a spin coherent state with $z$ being a complex time-dependent variable, $|IMK\rangle$ are the eigenstates of the intrinsic angular momentum operators $\hat{I}^{2}$ and $\hat{I}_{3}$ and their counterparts in the laboratory frame of reference, while $\hat{I}_{-}$ is a ladder operator. The averages on the variational state of the terms involved in the variation (\ref{var}) are calculated using the results of Refs.\cite{Radcliffe,Takeno,RaBu} and have the following expressions:
\begin{eqnarray}
\langle H_{chiral}\rangle&=&\frac{I}{2}(A_{1}+A_{2})+A_{3}I^{2}+\frac{I(2I-1)}{2(1+zz^{*})^{2}}\nonumber\\
&&\times\left[A_{1}(z+z^{*})^{2}-A_{2}(z-z^{*})^{2}-4A_{3}zz^{*}\right]-\nonumber\\
&&\frac{2A_{1}jI(z+z^{*})}{1+zz^{*}}+i\frac{2A_{2}j'I(z-z^{*})}{1+zz^{*}},\\
\left\langle\frac{\partial}{\partial{t}}\right\rangle&=&\frac{I(\dot{z}z^{*}-z\dot{z}^{*})}{1+zz^{*}}.
\end{eqnarray}
$z$ and its complex conjugate counterpart are considered as independent variables. The time dependent variational equation (\ref{var}) offers the following equations of motion for the complex variables $z$ and $z^{*}$:
\begin{equation}
\frac{\partial{\mathcal{H}}}{\partial{z}}=-\frac{2iI\dot{z}^{*}}{(1+zz^{*})^{2}},\,\,\,\,\frac{\partial{\mathcal{H}}}{\partial{z^{*}}}=\frac{2iI\dot{z}}{(1+zz^{*})^{2}},
\end{equation}
where $\mathcal{H}(z,z^{*})=\langle H_{chiral}\rangle$ plays now the role of a classical energy function which is also a constant of motion. For simplicity, the complex variable is written in a stereographic representation \cite{Radcliffe}
\begin{equation}
z=\tan{\frac{\theta}{2}}e^{i\varphi},\,\,\,0\leq\theta<\pi,\,\,\,0\leq\varphi<2\pi.
\end{equation}
Within this parametrization, the angular momentum carried by the coherent state is oriented in the direction specified by the two angles of rotation $\theta$ and $\varphi$ \cite{Takeno} as in Fig.\ref{elipsoid}. The equations of motion for the new variables are given as:
\begin{equation}
\frac{\partial{\mathcal{H}}}{\partial{\theta}}=-I\sin{\theta}\dot{\varphi},\,\,\,\,\frac{\partial{\mathcal{H}}}{\partial{\varphi}}=I\sin{\theta}\dot{\theta}.
\end{equation}

The full structure of the classical Hamiltonian system is reproduced if the variables are canonical. This is achieved by the change of variable
\begin{equation}
r=2I\cos^{2}{\frac{\theta}{2}},\,\,\,0< r\leq 2I.
\end{equation}
Note that this is not an unique choice, because $r$ is defined up to a constant. For example the obvious change of variable $r=I\cos{\theta}$ leads to the same canonical Hamilton form for the equations of motion:
\begin{equation}
\frac{\partial{\mathcal{H}}}{\partial{r}}=\dot{\varphi},\,\,\,\,\frac{\partial{\mathcal{H}}}{\partial{\varphi}}=-\dot{r}.
\label{motion}
\end{equation}
These equations identify $\varphi$ as the generalized coordinate, while $r$ as the generalized momentum. The two canonical variables are then related by the Poisson bracket
\begin{equation}
\{\varphi,r\}=1.
\label{poison}
\end{equation}
Within this notation, the equations of motion can be written as:
\begin{equation}
\{r,\mathcal{H}\}=\dot{r},\,\,\,\{\varphi,\mathcal{H}\}=\dot{\varphi}.
\end{equation}

The classical energy function have the following expression in terms of the canonical variables:
\begin{eqnarray}
\mathcal{H}(r,\varphi)&=&\frac{I}{2}(A_{1}+A_{2})+A_{3}I^{2}+\frac{(2I-1)r(2I-r)}{2I}\nonumber\\
&&\times(A_{1}\cos^{2}{\varphi}+A_{2}\sin^{2}{\varphi}-A_{3})-\nonumber\\
&&2A_{1}j\sqrt{r(2I-r)}\cos{\varphi}-2A_{2}j'\sqrt{r(2I-r)}\sin{\varphi}.\nonumber\\
\label{clase}
\end{eqnarray}
The conservation of the total angular momentum
\begin{equation}
I^{2}=I_{1}^{2}+I_{2}^{2}+I_{3}^{2},
\end{equation}
is guaranteed by the classical expressions of the angular momentum components as functions of the canonical variables \cite{RaBu,Ida}:
\begin{eqnarray}
I_{1}&=&\sqrt{r(2I-r)}\cos{\varphi},\nonumber\\
I_{2}&=&\sqrt{r(2I-r)}\sin{\varphi},\label{Icl}\\
I_{3}&=&r-I.\nonumber
\end{eqnarray}
In what follows, the $\varphi$ angle will be restricted to the interval $(0,90^{\circ})$, which corresponds to a situation when the total angular momentum and the single-particle spins share an octant of the three-dimensional space. This implies $\cos{\varphi}>0$ and $\sin{\varphi}>0$.

\setcounter{equation}{0}
\renewcommand{\theequation}{3.\arabic{equation}}
\section{Rotational dynamics}

The minimum points of the constant energy surface $\mathcal{H}(r,\varphi)=const.$ correspond to stable dynamical configurations. These are determined from:
\begin{eqnarray}
&&\left(\frac{\partial{\mathcal{H}}}{\partial{r}}\right)_{r_{0},\varphi_{0}}=0,\,\,\left(\frac{\partial{\mathcal{H}}}{\partial{\varphi}}\right)_{r_{0},\varphi_{0}}=0,\nonumber\\
&&Det\left[\left(\frac{\partial^{2}{\mathcal{H}}}{\partial{q_{i}}\partial{q_{j}}}\right)_{r_{0},\varphi_{0}}\right]>0,
\label{min}
\end{eqnarray}
where $i(j)=1,2$ with $q_{1}=r$ and $q_{2}=\varphi$.

At this point it is worth to recount that there are few possibilities in what concerns distribution of the total angular momentum on the principal axes of the intrinsic frame of reference \cite{FrauMeng}. When the total angular momentum lies within a principal plane, the situation is called planar. While an aplanar configuration designates a total angular momentum with non-vanishing projections on all principal axes. Solving thus the system of equations (\ref{min}), one obtains a critical point $(r_{p},\varphi_{p})$ corresponding to a planar case, with $r_{p}=I$ and $\varphi_{p}$ given as a solution of the equation
\begin{equation}
\frac{(2I-1)}{2}(A_{2}-A_{1})\cos{\varphi_{p}}\sin{\varphi_{p}}=A_{2}j'\cos{\varphi_{p}}-A_{1}j\sin{\varphi_{p}}.
\end{equation}
The planar nature of this critical point results from a vanishing third component of the total angular momentum for $r_{p}=I$. From the above equation one can see that $\varphi_{p}$ depends on $I$, except when $\gamma=90^{\circ}$ because then $A_{1}=A_{2}$. In this particular case one have just $\tan{\varphi_{p}}=j'/j$.

\begin{figure}[t!]
\begin{center}
\includegraphics[width=0.48\textwidth]{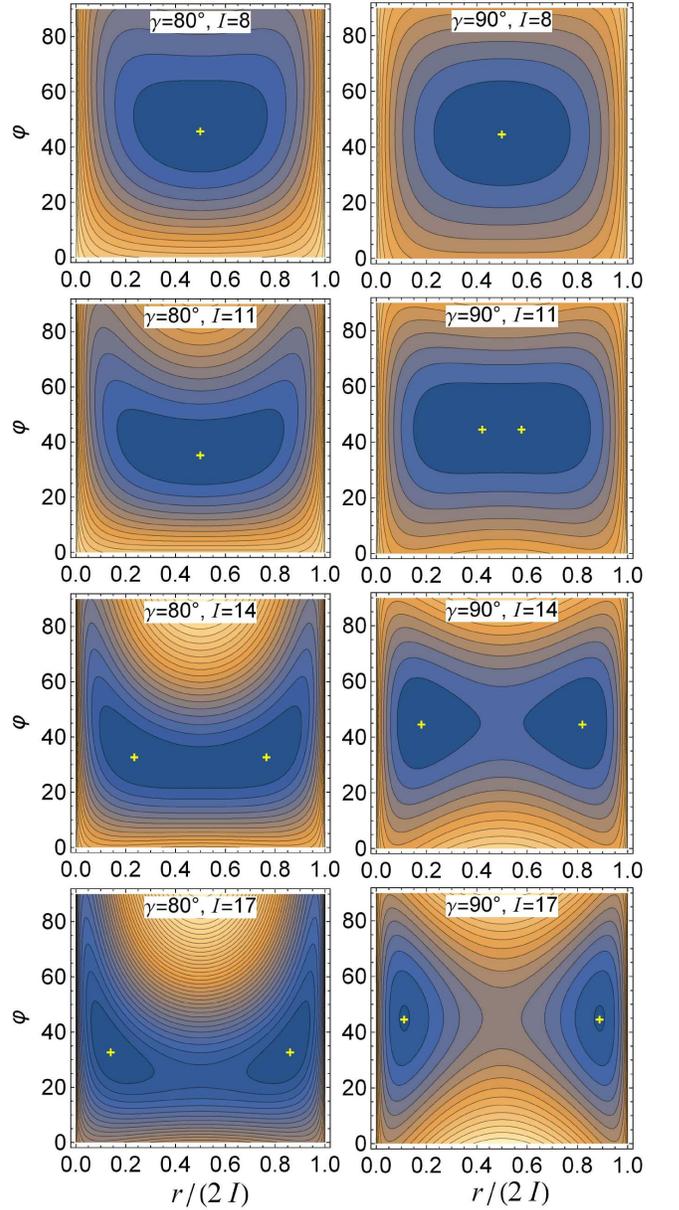}
\end{center}
\caption{Classical energy surfaces as a function of the generalized coordinate $\varphi$ and momentum $r$ for $\gamma=80^{\circ}$ and $\gamma=90^{\circ}$ and selected values of the total angular momentum. The single or double minima are indicated with crosses, while the difference between two consecutive contours is 10 arbitrary units. The increase goes from dark to light.}
\label{surf}
\end{figure}

\begin{figure}[t!]
\begin{center}
\includegraphics[width=0.48\textwidth]{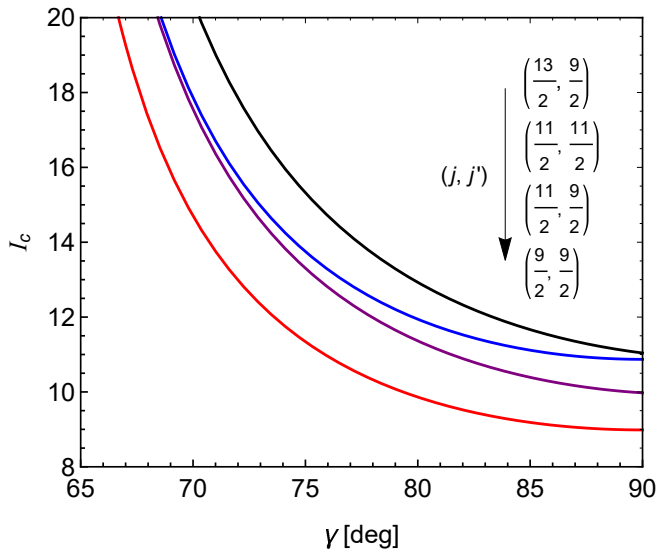}
\end{center}
\caption{The evolution as a function of triaxiality $\gamma$ of the separatrix represented by the critical angular momentum $I_{c}$ for few quasiparticle configurations expected to break the chiral symmetry.}
\label{Ic}
\end{figure}

Eqs. (\ref{min}) also provide an aplanar stationary point specified by:
\begin{eqnarray}
&&\sin{\varphi_{a}}=\frac{A_{2}j'(A_{1}-A_{3})}{\sqrt{A_{1}^{2}j^{2}(A_{2}-A_{3})^{2}+A_{2}^{2}j'^{2}(A_{1}-A_{3})^{2}}},\nonumber\\
&&\label{sinf}\\
&&\cos{\varphi_{a}}=\frac{A_{1}j(A_{2}-A_{3})}{\sqrt{A_{1}^{2}j^{2}(A_{2}-A_{3})^{2}+A_{2}^{2}j'^{2}(A_{1}-A_{3})^{2}}},\nonumber\\
\label{cosf}\\
&&\sqrt{r_{a}(2I-r_{a})}=I\sin{\theta_{a}^{I}},
\end{eqnarray}
where one used the following notation
\begin{equation}
\sin{\theta_{a}^{I}}=\frac{2\sqrt{A_{1}^{2}j^{2}(A_{2}-A_{3})^{2}+A_{2}^{2}j'^{2}(A_{1}-A_{3})^{2}}}{(2I-1)(A_{1}-A_{3})(A_{2}-A_{3})}.
\label{sint}
\end{equation}
The two solutions for $r_{a}$ will then be
\begin{eqnarray}
r_{R}&=&I(1+\cos{\theta_{a}^{I}}),\\
r_{L}&=&I(1-\cos{\theta_{a}^{I}}).
\end{eqnarray}
Note that angle $\theta_{a}^{I}$ is spin-dependent, while $\varphi_{a}$ is not. The indexes $R$ and $L$ denote the right-handed and respectively the left-handed total angular momentum orientation in respect to the intrinsic frame of reference. The right-handedness of a configuration is associated to the case when one can count in the mathematically positive direction the principal axes as 1, 2 and 3 when looking from the tip of the total angular momentum vector. This assignment of the two solutions is more obvious when the averages of the total angular momentum components are considered for this aplanar stationary point:
\begin{eqnarray}
I_{1}^{a}&=&I\sin{\theta_{a}^{I}}\cos{\varphi_{a}},\\
I_{2}^{a}&=&I\sin{\theta_{a}^{I}}\sin{\varphi_{a}},\\
I_{3}^{a}&=&\pm I\cos{\theta_{a}^{I}}.
\end{eqnarray}
This is just a representation of a vector of magnitude $I$ in spherical coordinates (see Fig.\ref{elipsoid}). The minimum condition for the aplanar solutions implies that the rational functions (\ref{sinf}), (\ref{cosf}) and (\ref{sint}) have under unity values. This is obvious for the first two, from their analytical expression. The latter however is more difficult to judge, but can be inferred from the successive changes of variables leading to the expression (\ref{sint}). Nevertheless, the condition $0<\sin{\theta_{a}^{I}}<1$ provides some additional restrictions on the relative distribution of the $A_{k}$ parameters which are also angular momentum dependent. As a matter of fact, the condition $\sin{\theta_{a}^{I}}=1$ serves as a separatrix which marks the border between the planar and aplanar solutions corresponding to two distinct rotational phases. This separatrix provides a critical angular momentum value at which the transition between the two phases commences from the planar phase to the aplanar one as is shown in the Fig.\ref{surf}. The critical angular momentum depends on the triaxiality measure $\gamma$ as well as the single-particle spins. Its evolution as a function of $\gamma$ is depicted in Fig.\ref{Ic} for few simple one particle one hole configurations commonly known to generate chiral symmetry breaking. The critical angular momentum value tends to infinity when the density distribution is axially symmetric, and has its minimum value at maximal triaxiality $\gamma=90^{\circ}$. The minimum values for the considered quasiparticle configurations are listed in Table \ref{tab1}.

\setlength{\tabcolsep}{5.5pt}
\begin{table}[ht!]
\caption{Minimal classical value of the critical angular momentum where the classical energy function starts to allow stable chiral solutions, for few one particle one hole quasiparticle configurations known to generate chiral doublet bands.}
\label{tab1}
\begin{center}
\begin{tabular}{cccccc}
\hline\hline\noalign{\smallskip}
Mass region&Configuration&$j$&$j'$&$I_{c}^{min}$\\
\noalign{\smallskip}\hline\noalign{\smallskip}
$A\sim80$& $\pi g_{\frac{9}{2}}\otimes\nu g_{\frac{9}{2}}^{-1}$ &$\frac{9}{2}$& $\frac{9}{2}$&8.99\\
\noalign{\smallskip}
$A\sim100$& $\pi g_{\frac{9}{2}}^{-1}\otimes\nu h_{\frac{11}{2}}$ &$\frac{11}{2}$& $\frac{9}{2}$&9.98\\
\noalign{\smallskip}
$A\sim130$& $\pi h_{\frac{11}{2}}\otimes\nu h_{\frac{11}{2}}^{-1}$ &$\frac{11}{2}$& $\frac{11}{2}$&10.87\\
\noalign{\smallskip}
$A\sim190$& $\pi h_{\frac{9}{2}}\otimes\nu i_{\frac{13}{2}}^{-1}$ &$\frac{13}{2}$& $\frac{9}{2}$&11.04\\
\noalign{\smallskip}\hline\hline
\end{tabular}
\end{center}
\vspace{-0.5cm}
\end{table}

From Fig.\ref{surf} one observes that in the planar phase, the classical energy function has a single minimum at $r_{p}=I$ and $\varphi_{p}$, which becomes a saddle point after crossing the separatrix. In turn, the saddle point marks the apparition of the two chiral minima at $(r_{R},\varphi_{a})$ and $(r_{L},\varphi_{a})$. Although using different variables, the energy surfaces of Fig.\ref{surf} are similar to the total Routhian surface calculations made in \cite{col1} considering a single orientation angle, especially when $\gamma\neq90^{\circ}$. The connection to azimuthal and polar angles can be easily made, one however maintained the $(r,\varphi)$ space because the two variables are canonical conjugate. The major difference arises for the maximal triaxiality case ($\gamma=90^{\circ}$), where the classical energy function is doubly symmetric in respect to $\varphi=45^{\circ}$ and $r=I(I_{3}=0)$ lines. This two-fold symmetry is however recovered when the total Routhian is considered in the full space of the two orientation angles \cite{col2}.

\begin{figure}[t!]
\begin{center}
\includegraphics[width=0.48\textwidth]{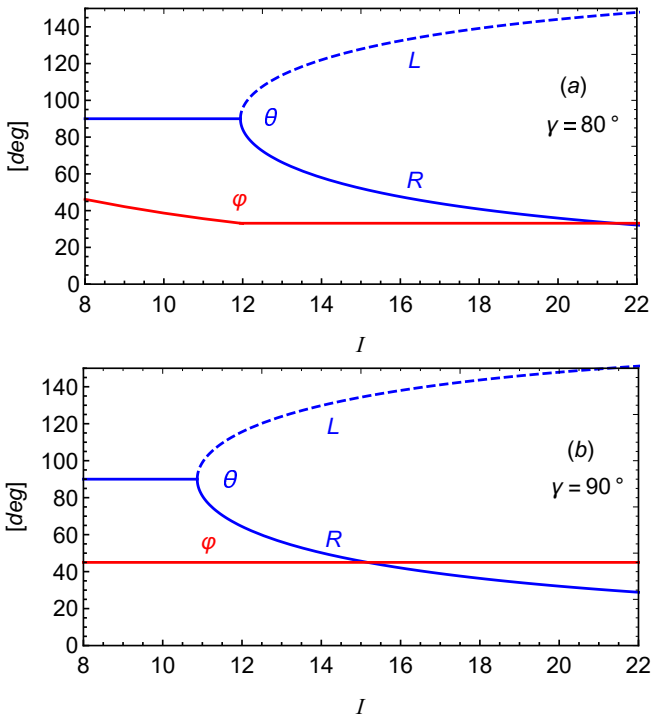}
\end{center}
\caption{Evolution of the spherical angles as a function of angular momentum for different degrees of triaxiality: (a) $\gamma=80^{\circ}$ and (b) $\gamma=90^{\circ}$.}
\label{angles}
\end{figure}

\begin{figure}[t!]
\begin{center}
\includegraphics[width=0.48\textwidth]{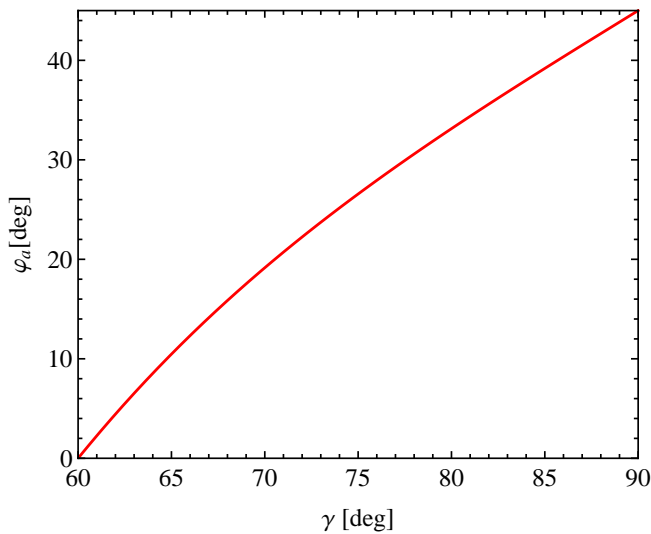}
\end{center}
\caption{The correspondence between the triaxiality measure $\gamma$ and the final value of the azimuthal angle $\varphi_{a}$, which remains invariant with total angular momentum.}
\label{faga}
\end{figure}

At this point, one can analyze the dynamical evolution, {\it i.e.} as a function of total angular momentum modulus, of the tilting angles defining the average geometrical direction of the total angular momentum vector. This is best presented graphically in Fig.\ref{angles}, where one plotted the polar and azimuthal angles as function of the total angular momentum for few asymmetrical values of $\gamma$. In consensus with the previous observations, the starting value of the polar angle $\theta$ is $90^{\circ}$. This value persists throughout the entire planar phase up to the critical value $I_{c}$, where it bifurcates into the two chiral branches with $\theta=\theta_{a}^{I}$ and $\theta=\pi-\theta_{a}^{I}$. The existence of the planar phase at small angular momentum values is due to the sizable components of the core angular momentum on the principal axes 1 and 2 \cite{Shi} which add up to the single particle contributions. The planar average direction of the total spin is however soft against out of plane fluctuations as can be attested by the pronounced shallowness of the planar minima. In what concerns the azimuthal angle $\varphi$, it has an invariant value of $45^{\circ}$ for maximal triaxiality $\gamma=90^{\circ}$, while for $\gamma\neq90^{\circ}$ it just starts from this value and is continuously decreasing up to the critical point keeping the corresponding tilting constant through the evolution in the aplanar phase. The correspondence between the constant value of the azimuthal angle acquired at the critical point and the triaxiality degree is visualized in Fig.\ref{faga}.

\setcounter{equation}{0}
\renewcommand{\theequation}{4.\arabic{equation}}
\section{Emergence of chiral bands}

As the energy function has always a single minimum only in the $\varphi$ variable, one chooses to expand it around the corresponding minimum points for fixed values of $r$:
\begin{equation}
\tilde{\mathcal{H}}(r,\varphi)\approx\mathcal{H}(r,\varphi_{0}(r))+\frac{1}{2}\left(\frac{\partial^{2}{\mathcal{H}}}{\partial{\varphi}^{2}}\right)_{\varphi_{0}(r)}\tilde{\varphi}^{2},
\label{Haprox}
\end{equation}
where $\tilde{\varphi}=\varphi-\varphi_{0}(r)$ with $\varphi_{0}(r)$ being the value which minimizes the energy function for a fixed $r$. $\varphi_{0}(r)$ is therefore defined as the solution of the following equation
\begin{eqnarray}
&&(2I-1)r(2I-r)(A_{2}-A_{1})\cos{\varphi_{0}}\sin{\varphi_{0}}\nonumber\\
&&=2I\sqrt{r(2I-r)}\left(A_{2}j'\cos{\varphi_{0}}-A_{1}j\sin{\varphi_{0}}\right).
\end{eqnarray}
Lacking an analytical expression for the general solution of the above equation, one will further pursue only the special case of $\gamma=90^{\circ}$, for which $A_{1}=A_{2}$ and $\varphi_{0}=45^{\circ}=const.$ The general case for $\gamma\neq90^{\circ}$ implies a numerical part and will be presented elsewhere.

In order to have a better view of the chiral dynamics, a new chiral variable is introduced, namely $\tilde{r}=r-I$. This quantity is just the classical third component of the total angular momentum (\ref{Icl}), and varies between $-I$ and $I$. The pair of variables $\tilde{r}$ and $\tilde{\varphi}$ are also canonical conjugate, {\it i.e.} $\{\tilde{r},\tilde{\varphi}\}=1$. Symmetrizing the products of $\tilde{r}$ and $\tilde{\varphi}$ functions one proceeds to the quantization of the approximate classical energy function (\ref{Haprox}) by making the substitutions
\begin{equation}
\tilde{r}=x,\,\,\,\tilde{\varphi}=i\frac{d}{dx},
\end{equation}
rather than quantizing the classical trajectories by means of a WKB-like approximation as was performed in Ref.\cite{Malik}. This differential representation of the conjugate canonical coordinates is equivalent to working in the momentum space represented by the generalized momentum variable $\tilde{r}$. After the quantization procedure, one arrives at a quantum Hamiltonian expressed as the differential operator
\begin{eqnarray}
\hat{H}_{c}&=&-\frac{1}{2B(x)}\frac{d^{2}}{dx^{2}}+\frac{B'(x)}{2\left[B(x)\right]^{2}}\frac{d}{dx}+\mathcal{H}(x,\varphi_{0})+\nonumber\\
&&\frac{B''(x)}{4\left[B(x)\right]^{2}}-\frac{\left[B'(x)\right]^{2}}{2\left[B(x)\right]^{3}},
\label{ecdiag}
\end{eqnarray}
where
\begin{equation}
B(x)=\left[\frac{\partial^{2}{\mathcal{H}(x,\varphi)}}{\partial{\varphi}^{2}}\right]_{\varphi_{0}}^{-1}.
\end{equation}

\begin{figure}[t!]
\begin{center}
\includegraphics[width=0.48\textwidth]{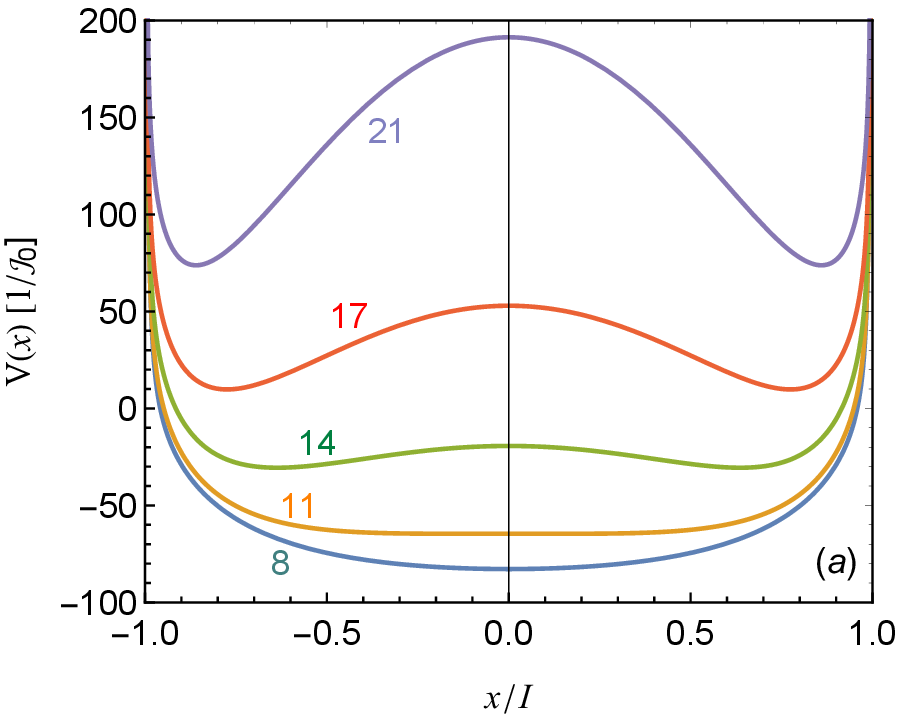}\\
\includegraphics[width=0.48\textwidth]{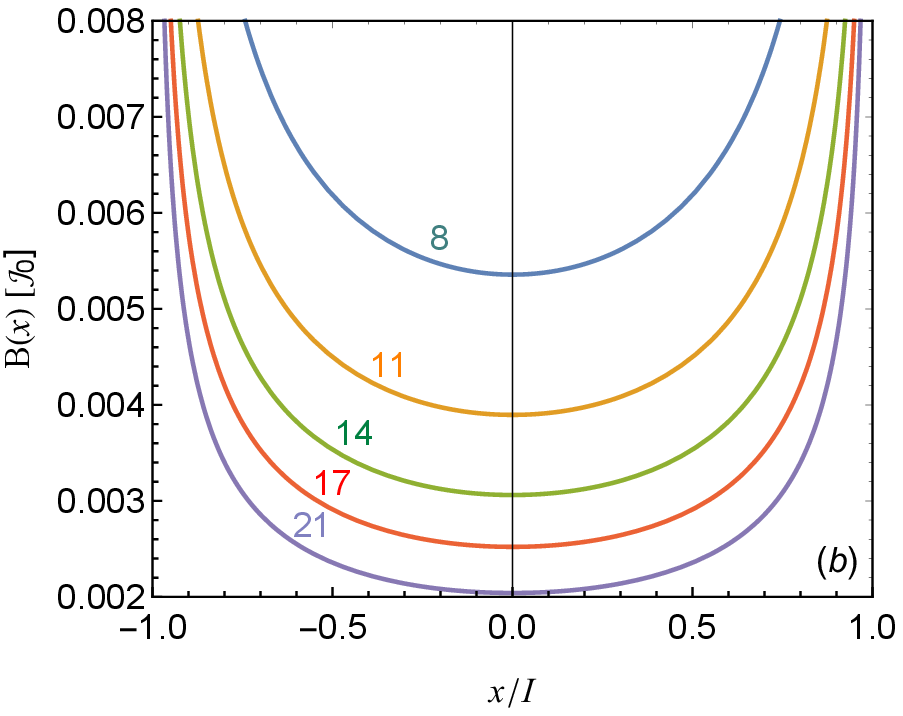}
\end{center}
\caption{Chiral potential (a) and mass (b) as function of the chiral variable $x$ for few values of total angular momentum.}
\label{potmass}
\end{figure}

Suppose now that the wave function corresponding to above quantum Hamiltonian is $F(x)$ and is normalized to unity. Then making the change of function $f(x)=[B(x)]^{-1/4}F(x)$, one can write the final Hamiltonian for $f(x)$ in the following form
\begin{equation}
\hat{H}_{c}=-\frac{1}{2}\frac{1}{\sqrt{B(x)}}\frac{d}{dx}\frac{1}{\sqrt{B(x)}}\frac{d}{dx}+V(x),
\label{ecx}
\end{equation}
where $B(x)$ plays the role of an one dimensional mass which depends on $x$, while the corresponding potential is given as:
\begin{equation}
V(x)=\mathcal{H}(x,\varphi_{0})+\frac{B''(x)}{8\left[B(x)\right]^{2}}-\frac{9\left[B'(x)\right]^{2}}{32\left[B(x)\right]^{3}}.
\end{equation}
The identification of $B(x)$ as the mass of the system is confirmed also by the normalization condition for the function $f(x)$ which reads as
\begin{equation}
\int f(x)f^{*}(x)\sqrt{B(x)}dx=1.
\end{equation}
It can be easily checked that both mass function and the potential are invariant under the parity transformation $x\rightarrow-x$. All these ingredients are reminiscent of the one dimensional collective Hamiltonian obtained in \cite{col1}. The difference here is that both mass term and potential are products of the original quantum Hamiltonian, and are determined solely on the basis of the rotational geometry, in comparison to the approach of Ref.\cite{col1} where the kinetic and potential terms are obtained in separate ways. Moreover, due to the canonical conjugate character of the two semiclassical coordinates $r$ and $\varphi$, there is an additional relation between the polar and azimuthal angles. Thus, although the chiral Hamiltonian is one-dimensional, the quantum fluctuations of both directional angles are included. From the graphical representation of the chiral potential and mass term shown in Fig.\ref{potmass}, one can see that while the emergence of the double minimum profile for the chiral potential is similar to that of Ref.\cite{col1}, the dynamical evolution of the mass is quite different. Indeed, the mass term determined in Ref.\cite{col1} starts from being shallow at low rotational frequencies acquiring a more localized minimum as the frequency is increased. In the present case, the evolution is opposite as can be seen from Fig.\ref{potmass}(b). This distinction comes from the fact that in the present case, the mass is defined in the momentum space. Otherwise the picture is consistent with the formalism of Ref.\cite{col1}.

The states of the two chiral partner bands are then defined by the first two eigensolutions of the differential operator (\ref{ecdiag}). Although the associated mass term and the chiral potential have analytical expressions, the corresponding Schr\"{o}dinger equation cannot be exactly solved. Consequently, the energies are determined through a diagonalization in a suitable basis. In order to avoid large dimension diagonalizations, it is customary to use different basis states for even and odd parity solutions when symmetric potentials are involved. Choosing particle in the box eigenstates as basis functions, one assigns for even parity the basis states
\begin{equation}
g_{n}^{1}(x)=\frac{1}{\sqrt{I}}\cos{\left[\frac{(2n-1)\pi x}{2I}\right]},\,\,n=1,2,...,
\end{equation}
while for the odd parity the following basis states are used:
\begin{equation}
g_{n}^{-1}(x)=\frac{1}{\sqrt{I}}\sin{\left[\frac{2n\pi x}{2I}\right]},\,\,n=1,2,...
\end{equation}
These functions, like the exact eigenfunctions, satisfy the Dirichlet boundary condition
\begin{equation}
g_{n}^{p}(I)=g_{n}^{p}(-I)=0,
\end{equation}
and where shown to be very performant as basis states in symmetrical multiple minima problems \cite{Taseli,Chandra}. The eigenvalues of Eq.(\ref{ecx}) are then obtained by diagonalization in the above defined basis space which is truncated such that to accommodate a satisfactory convergence of the results. The same procedure will give the coefficients $a_{n}$ of the basis expansion
\begin{equation}
F_{p}(x)=\sum_{n=1}^{N}a_{n}^{p}g_{n}^{p}(x),\,\,p=-1,1,
\end{equation}
where $N$ denotes the dimension of the truncated space.

\begin{figure}[t!]
\begin{center}
\includegraphics[width=0.47\textwidth]{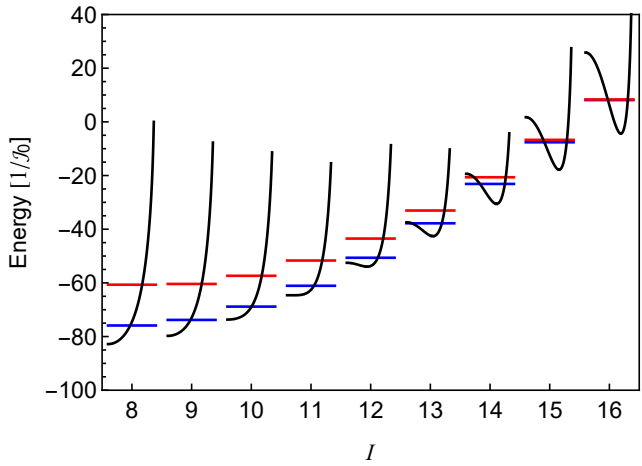}
\end{center}
\caption{The lowest two eigenvalues of the chiral potential for $I=8-16$ are visualized relative to the potential profile. The lowest energy state corresponds to the symmetric wave function ($p=1$).}
\label{diag}
\end{figure}

The splitting between energies of the two chiral solutions is shown in Fig.\ref{diag} relative to the barrier hight and the depth of the minimum for a series of integer values of total angular momentum. The splitting persists along many angular momentum states, vanishing only when the two energy states become considerably lower than the barrier peak, that is around $I=15$. The results are consistent with the well known behaviour of the spectra for double well potentials \cite{Pitzer}.

\setcounter{equation}{0}
\renewcommand{\theequation}{5.\arabic{equation}}
\section{Total wave-functions and electromagnetic transitions}

\begin{figure}[t!]
\begin{center}
\includegraphics[width=0.47\textwidth]{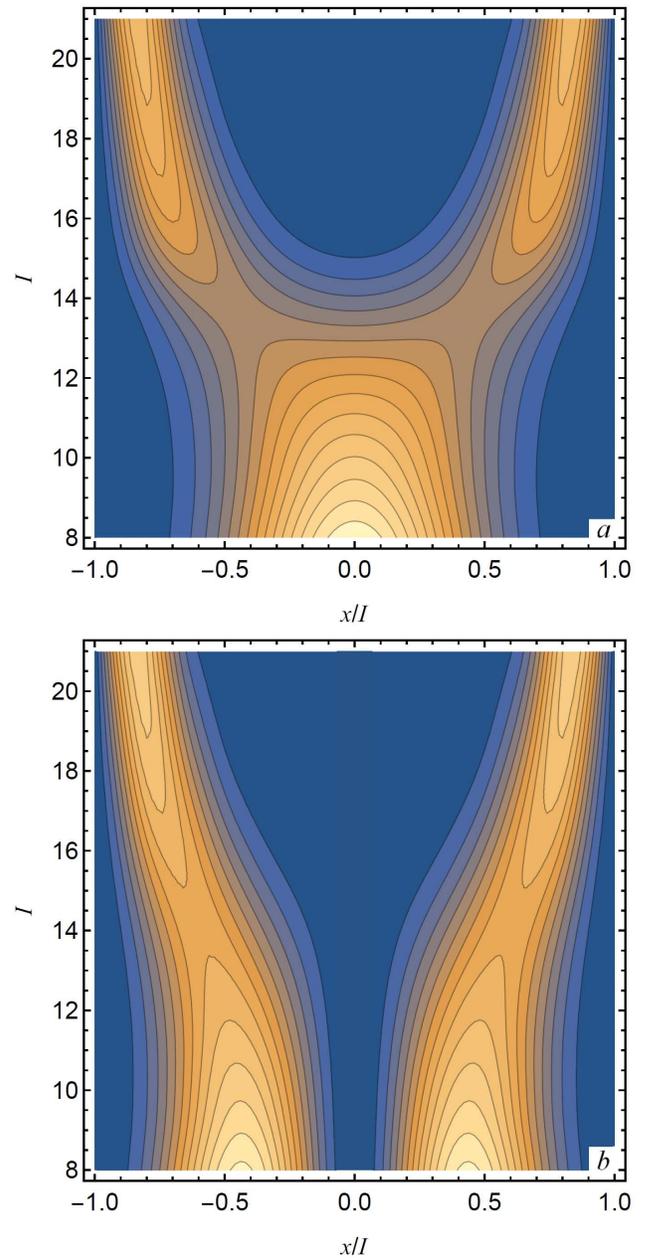}
\end{center}
\caption{Density probability distribution as function of total angular momentum and the chiral variable $x=K$ for ground (a) and first excited (b) states corresponding to $p=1$ and respectively $p=-1$. Consecutive contours denote a variation of probability of 0.01 arbitrary units. The increase goes from dark to light.}
\label{dens}
\end{figure}

Expressing the original complex variable in terms of the chiral one as
\begin{equation}
z=\sqrt{\frac{I-x}{I+x}}e^{i\varphi},
\end{equation}
one can right down the coherent state (\ref{coh}) as
\begin{eqnarray}
|\psi(x,\phi)\rangle&=&\sum_{K=-I}^{I}\frac{1}{(2I)^{I}}\sqrt{\frac{(2I)!}{(I-K)!(I+K)!}}\nonumber\\
&&\times(I+x)^{\frac{I-K}{2}}(I-x)^{\frac{I+K}{2}}\nonumber\\
&&\times e^{i\varphi(I+K)}|IMK\rangle.
\end{eqnarray}
In order to couple the rotational motion described by the above state with the information regarding the chiral vibration, one will weight the coherent state in $\varphi=\varphi_{0}=45^{\circ}$ with the density probability for the oscillating chiral variable
\begin{equation}
\rho^{I}_{p}(x)=\left|F^{I}_{p}(x)\right|^{2},\,\,p=-1,1.
\end{equation}
The evolution of this quantity with total angular momentum can be tracked in Fig.\ref{dens}, where one plotted the interpolated density probability as function of $x$ and $I$.

The wave functions with restored chiral symmetry can be then expressed as:
\begin{equation}
|IMp\rangle=\mathcal{N}_{Ip}\sum_{K=-I}^{I}S_{IKp}e^{i\varphi_{0}(I+K)}|IMK\rangle,
\end{equation}
where
\begin{eqnarray}
S_{IKp}&=&\frac{1}{(2I)^{I}}\sqrt{\frac{(2I)!}{(I-K)!(I+K)!}}\\
&&\times\int_{-I}^{I}\rho^{I}_{p}(x)(I+x)^{\frac{I-K}{2}}(I-x)^{\frac{I+K}{2}}dx,\nonumber
\end{eqnarray}
while $\mathcal{N}_{Ip}$ is a redefined normalization constant. Using these wave functions, one can now proceed to the calculation of the quadrupole transition probabilities using the following transition operator:
\begin{equation}
\mathcal{M}(E2,\mu)=\sqrt{\frac{5}{16\pi}}\left[Q'_{0}D_{\mu0}^{2}+\frac{Q'_{2}}{\sqrt{2}}\left(D_{\mu2}^{2}+D_{\mu-2}^{2}\right)\right].
\end{equation}
$Q'_{0}$ and $Q'_{2}$ are intrinsic quadrupole moments for a reference frame where the MOI on the third principal axis is maximal. They can be related to the commonly used components $Q_{0}=Q\cos{\gamma}$ and $Q_{2}=Q\sin{\gamma}/\sqrt{2}$ defined in a system of reference with the maximal MOI along the first axis, by
\begin{eqnarray}
Q'_{0}&=&-\frac{1}{2}Q_{0}+\sqrt{\frac{3}{2}}Q_{2}=-Q\cos{\left(\gamma+\frac{\pi}{3}\right)},\\
Q'_{2}&=&-\frac{1}{2}\left(\sqrt{\frac{3}{2}}Q_{0}+Q_{2}\right)=-Q\frac{\sin{\left(\gamma+\frac{\pi}{3}\right)}}{\sqrt{2}},
\end{eqnarray}
where $Q=\frac{3}{\sqrt{5\pi}}R_{0}^{2}Z\beta$ with $\beta$ being the axial deformation, $Z$ is the charge number, while $R_{0}$ is the nuclear radius.

The reduced transition probability is determined with
\begin{equation}
B(E2,Ip\rightarrow I'p')=\left|\langle Ip||\mathcal{M}(E2)||I'p'\rangle\right|^{2}.
\end{equation}
The expression for the involved reduced matrix element of the quadrupole transition operator in the considered particular case of $\gamma=90^{\circ}$ and $\varphi_{0}=45^{\circ}$ can be readily deduced:
\begin{eqnarray}
&&\langle Ip||\mathcal{M}(E2)||I'p'\rangle=\frac{Q}{8}\sqrt{\frac{15}{\pi}}\frac{\hat{I'}}{\hat{I}}e^{i\frac{\pi}{4}(I'-I)}\\
&&\times\sum_{K=-I}^{I}S_{IKp}S_{I'Kp'}C^{I'\,2\,I}_{K\,0\,K}\nonumber.
\end{eqnarray}
The simple form is obtained by dismissing the non-diagonal quadrupole components which cancel each other when the summation is performed on positive and negative projections.

Another observable related to chiral partner bands, is the magnetic dipole transition probability \cite{PRM2,PRM7,Petrache,Tonev2006,Tonev2007,Qi}. It is however predominantly given by the single-particle degrees of freedom which are neglected in the present study.

\renewcommand{\theequation}{6.\arabic{equation}}
\section{Comparison with experiment}

\begin{figure}[t!]
\begin{center}
\includegraphics[width=0.47\textwidth]{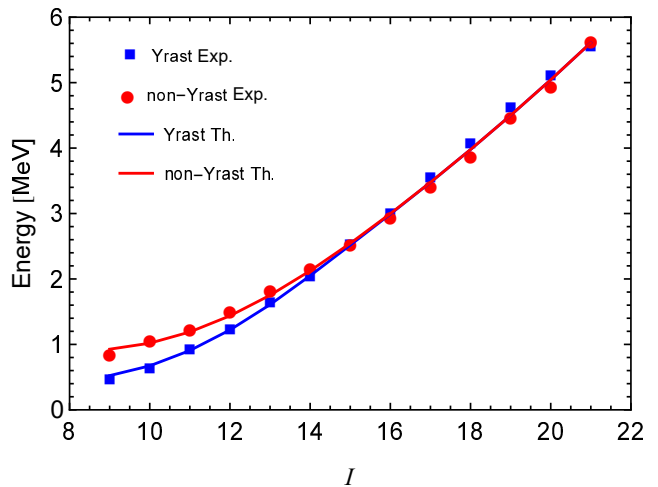}
\end{center}
\caption{Comparison of yrast and non-yrast energy levels between theoretical results and experimental data \cite{Tonev2007} for $^{134}$Pr.}
\label{expe}
\end{figure}

The formalism is applied to the chiral bands of $^{134}$Pr, which are among the most extended in what concerns the number of observed different spin states. For the calculation of the energy levels corresponding to the two partner bands the following formula is used:
\begin{equation}
E_{Ip}=E_{0}+E^{c}_{Ip},
\end{equation}
where $E_{0}$ is an energy reference, while $E_{Ip}^{c}$ is the eigenvalue of the chiral quantum Hamiltonian, obtained from the diagonalization procedure. The dimension of the diagonalization basis is truncated at 50 states, assuring thus a convergence of the energies up to spin $I=21$. As the asymmetry of the triaxial core $\gamma$ is considered fixed at $90^{\circ}$, the only free parameters remain the reference energy $E_{0}$ and the inertial constant $\mathcal{J}_{0}$. Fitting the experimental data against the two parameters one obtains the following values: $E_{0}=2.746$ MeV, $\mathcal{J}_{0}=33.196$ MeV$^{-1}$, which correspond to an $rms$ of 59.28 keV. The exceptionally good reproduction of data can be better seen in Fig.\ref{expe}, where all aspects of the data evolution, such as the general rotational behaviour of the two bands, energy splitting between bands, as well as the angular momentum of critical point for the transition between chiral vibration and static chirality, are well reproduced.

\begin{figure}[t!]
\begin{center}
\includegraphics[width=0.47\textwidth]{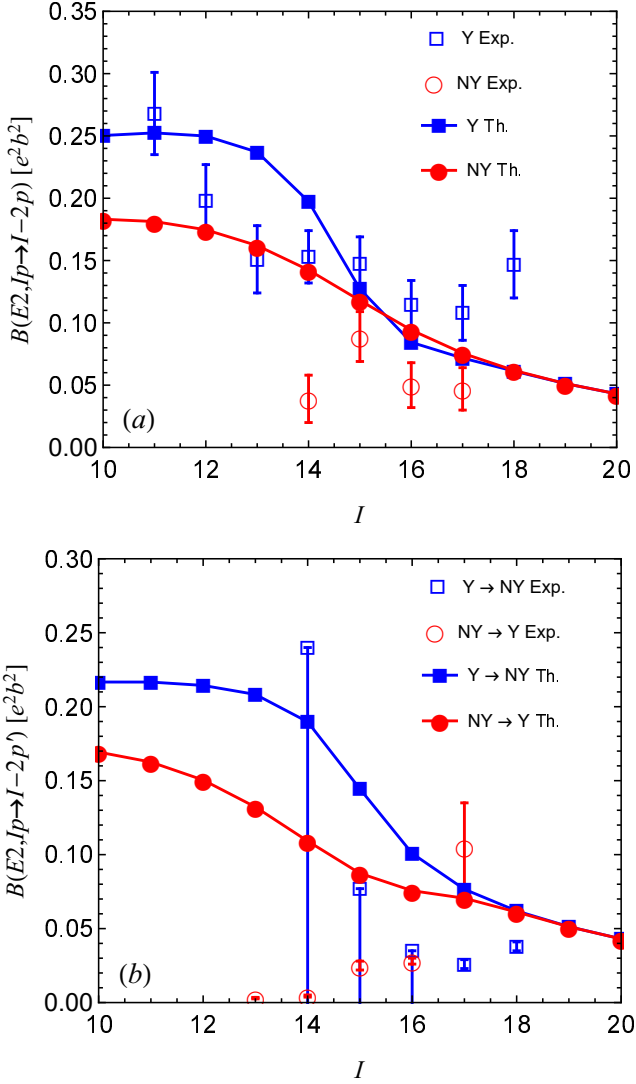}
\end{center}
\caption{Theoretical values of $B(E2)$ are compared to the experimental data measured for $^{134}$Pr for intra-band (a) and inter-band (b) transitions involving yrast (Y) and non-yrast (NY) states.}
\label{tranz}
\end{figure}

For the calculation of $E2$ transition probabilities, the value $Q=3.5\, e\,b$ is considered as in Refs.\cite{PRM5,PRM7,Qi}. The theoretical results are compared to experimentally available data on $^{134}$Pr in Fig.\ref{tranz}. The agreement with experiment is satisfactory, with a better reproduction of the data for intra-band transitions. Especially well reproduced is the descending trend of intra-band transitions for $I=15-17$. The evolution of theoretical results with angular momentum is similar for both intra- and inter-band transitions. In both cases, the transitions from yrast states are, with few exceptions, greater than those from the non-yrast states up to $I=16$. The same is true for the measured values. For $I\geq17$, the two transition probabilities become equal due to the stabilization of the static chirality. The difference between $B(E2)$ from yrast and those from non-yrast states is almost constant up to $I=13$. Starting form this angular momentum value, all transition rates undergo a kind of second order phase transition to lower values \cite{Landau}. The difference becomes first larger and then smaller for the intra-band transitions, while the inter-band ones become continuously closer intersecting each other between $I=15$ and $I=16$.

The transitional region $I=14-16$ coincides with the angular momentum interval where the density probability of the vibrational states is most extended. Indeed, although from the semiclassical analysis, the critical angular momentum where chiral minima appear in the classical energy function is $I=11$ (Table \ref{tab1}), from quantum point of view the two chiral solutions become distinguishable only around $I=16$ where the quantum tunneling subsides. Fig.\ref{tranz} shows that transition probabilities involving the state $I=14$ act as critical points for the change from high to low $B(E2)$ values. Coming back to the density probability distribution depicted in Fig.\ref{dens}, one can see that the density probability for $I=14$ in ground state covers both chiral minima with undistinguishable peaks, while for the excited state, the height of the two vibrational peaks is minimal. In the first case there exists a coexistence between the two chiral solutions. As a matter of fact the broadening of the probability density distribution attributed to coexistence phenomena have immediate repercussions on the electromagnetic properties \cite{Noi1,Noi2,Noi3}.

The good agreement with experimental energy levels, and electromagnetic transitions at least for a small interval of angular momentum states, indicates that chiral geometry is a viable hypothesis in what concerns the interpretation of the doublet bands observed in increasingly more nuclei. There are however alternative interpretations of the fingerprints usually attributed to nuclear chirality \cite{Meng1}. For example the interacting boson-fermion-fermion model analysis made on partner bands of $^{134}$Pr point to the domination of shape fluctuations over the chiral geometry \cite{Tonev2006,Tonev2007}. Among the alternative mechanisms of the doublet bands generation in $^{134}$Pr, one must mention also the shape coexisting scenario \cite{Petrache} where the two bands are considered to have different quadrupole moments. Therefore the chiral symmetry breaking cannot be considered the unique or the sole mechanism responsible for the experimentally observed doublet bands.

\section{Conclusions}

Through a time dependent variational principle one associated a classical energy function to a system of three mutually perpendicular spins corresponding to a triaxial core and two single particle configurations of valence nucleons. A coherent state for the angular momentum operators is used as a variational state, whose stereographic parametrization, gives the dependence of the classical energy function on azimuthal angle $\varphi$ and a canonical conjugate coordinate $r$ related to the polar angle $\theta$. Maintaining rigid the trihedral configuration of the three spins, it is found that the classical energy function goes from a single minimum to a double minima surface in the space of canonical variables $(\varphi,r)$ as the total angular momentum is increased. The analytical expression for these critical points identifies the solution with a single minimum as planar, while the double minimum solution is associated to an aplanar case. The two degenerated minima in the later case describe distinct chiral configurations of the three spin vectors involved in the dynamics of the total system.

The single minimum and double minima conditions define two distinct rotational phases which are delimited by a separatrix represented by a critical angular momentum value. The dependence of the critical spin on traiaxiality $\gamma$ for different single-particle configurations revealed that its minimum lies at maximum triaxiality $\gamma=90^{\circ}$. By studying the evolution with total angular momentum of the spherical angles associated to energy minima, a distinct dynamical behaviour was observed for $\gamma=90^{\circ}$. Speculating the symmetry of the classical energy function for this particular case, one quantized the energy function by replacing some redefined canonical conjugate coordinates with their corresponding differential operators after performing a harmonic approximation against one of the original coordinates. The resulting differential operator is written in terms of a new variable which is just the total angular momentum projection on the quantization axis. It was shown that the differential equation can be brought to a Schr\"{o}dinger form containing a kinetic operator with a variable-dependent mass term and an effective symmetrical potential which can have a single or double degenerated minima, depending on the total angular momentum.

The energy states of the chiral partner bands for a given angular momentum are obtained through diagonalization of the quantum Hamiltonian in a trigonometric basis with symmetric and antisymmetric basis states. The solutions are then used to calculate $B(E2)$ transition probabilities with a redefined total wave-function having an incorporated coupling between the rotational motion and chiral vibration. The model was applied to the description of the chiral bands of $^{134}$Pr. The agreement with experiment is very good in what concerns the energy levels considering that the triaxiality is a priori fixed to $\gamma=90^{\circ}$. Although the single-particle degrees of freedom are ignored because one considered rigid alignments of the single-particle spins, the agreement between theoretical calculations for the transition probabilities and experimental data is quite satisfactory. Especially good closeness to data is obtained for the transitional interval of angular momenta defining the change from chiral vibration to static chirality.

Although the considered system is drastically restrained, it provides a good reference picture for how the chiral symmetry breaking occurs and how it affects the system's rotation. The rotational aspect is mainly given by the classical analysis which sorts the relevant degrees of freedom further used to quantize the fluctuations around or between stable rotational configurations. Therefore, the proposed semiclassical approach is able to describe consistently the complex dynamics of a nucleus undergoing a transition from chiral vibration to static chirality.

\section*{Acknowledgments}
The author is grateful to Dr. Q. B. Chen for inspiring discussions. This work was supported by a grant of Ministry of Research and Innovation, CNCS - UEFISCDI, project number PN-III-P1-1.1-TE-2016-0268, within PNCDI III.

\end{document}